\documentclass[10pt,prl,superscriptaddress,preprintnumbers,amsmath,amssymb,nofootinbib]{revtex4}
\usepackage{epsfig}
\usepackage{amsmath}
\usepackage{graphicx}
\usepackage{amssymb}
\usepackage{bm}
\usepackage{dcolumn}
\usepackage{color}

\newcommand{\beq}{\begin{equation}}
\newcommand{\eeq}{\end{equation}}
\newcommand{\ba}{\begin{array}}
\newcommand{\ea}{\end{array}}
\newcommand{\bea}{\begin{eqnarray}}
\newcommand{\eea}{\end{eqnarray}}
\def\nn{\nonumber}

\begin{document}

\preprint{}

\title{Lepton flavor non-universality in the B-sector: a global analyses of various new physics models}

\author{Ashutosh Kumar Alok\footnote{akalok@iitj.ac.in}}
\affiliation{Indian Institute of Technology Jodhpur, Jodhpur 342037, India}

\author{Jacky Kumar\footnote{singlajacky@gmail.com}}
\affiliation{Department of High Energy Physics, Tata Institute of Fundamental Research, 400005, Mumbai, India}

\author{Dinesh Kumar\footnote{dinesh@uniraj.ac.in}}
\affiliation{Department of Physics, University of Rajasthan, Jaipur 302004, India}

\author{Ruchi Sharma\footnote{sharma.16@iitj.ac.in}}
\affiliation{Indian Institute of Technology Jodhpur, Jodhpur 342037, India}

\begin{abstract}
The measurements of the ratios $R_{K^{(*)}}$  along with $R_{D^{(*)}}$ hint towards lepton flavor non universality which is in disagreement with the standard model. In this work, we reanalyze the four new physics models, which are widely studied in the literature as a candidates for the simultaneous explanations of these measurements. These are, standard model like vector boson (VB), $SU(2)_L$-singlet vector leptoquark ($U_1$),  $SU(2)_L$-triplet scalar leptoquark ($S_3$) and  $SU(2)_L$ triplet vector leptoquark ($U_3$) models. We assume a coupling only to the  third generation in the weak basis, so that the $b \to s \mu^+ \mu^-$ transition is generated only via mixing effects. Preforming a global fit to all relevant data, we show that the vector boson model violates the current upper bound on ${Br}(\tau \to 3\mu)$ and hence is {\it inconsistent} with the present data. Further, we show that within this framework, the $U_1$ leptoquark model {\it cannot} simultaneously accommodate $R_{K^{(*)}}$ and $R_{D^{(*)}}$ measurements. We emphasize that this conclusion is independent of the additional constraints coming from renormaliztion group running effects and high-$p_T$ searches. In addition, we show that the $S_3$ and $U_3$ models are highly disfavored by the constraints coming from $b\to s \nu \bar \nu$ data. Finally, we find a that hypothesis of two LQ particles is also challenged by $b\to s \bar \nu \nu$ data. 
\end{abstract}


\maketitle
\section{\bf Introduction}

Apart from confirming some of the prevailing anomalies 
in the $B$-sector, the currently running LHC has provided 
several new measurements which hint towards physics beyond 
standard model (SM). Some  of these measurements are 
indicating towards lepton universality violation.  
The most striking measurements hinting towards lepton 
flavor non universality are the  $R_{D^{(*)}}\equiv \Gamma(B \to D^{(*)}\, \tau \bar{\nu})/\Gamma(B \to D^{(*)}\, l \bar{\nu})$ $(l=e,\,\mu)$ \cite{Lees:2012xj,Lees:2013uzd,Huschle:2015rga,Aaij:2015yra,Sato:2016svk,Hirose:2016wfn,Aaij:2017uff,Hirose:2017dxl,Aaij:2017deq} which disagree with the SM at $\sim$ 3.8$\sigma$ level \cite{average}.  In Moriond 2019, the Belle collaboration has announced new preliminary measurements of $R_{D^{(*)}}$ using semileponic tag \cite{rdrdstarnew}. These measurements are consistent with the SM at the level of 1.2$\sigma$. With the addition of these new measurements, the tension between $R_{D^{(*)}}$ experimental  world average and the SM prediction reduces to 3.1$\sigma$.

Lepton flavor universality violation was further corroborated  by the measurement of $R_K \equiv  \Gamma(B^+ \to K^+ \,\mu^+\,\mu^-)/\Gamma(B^+ \to K^+\,e^+\,e^-)$ = $ 0.745^{+0.090}_{-0.074}~{\rm (stat)} \pm 0.036~{\rm (syst)}$ \cite{rk}. This measurement was performed in the low dilepton invariant mass-squared $q^2$ range ($1.0 \le q^2 \le 6.0 \, {\rm GeV}^2$) and it deviates from the SM prediction, which is $\simeq 1$ \cite{Hiller:2003js, Bordone:2016gaq}, by 2.6$\sigma$.  This measurement has been recently updated in Moriond 2019. The updated value of 
$R_K$ is $0.846^{+0.060+0.016}_{-0.054-0.014}$~\cite{Aaij:2019wad} which is $\sim 2.5\sigma$ away from the SM.
In April 2018, the LHCb collaboration  announced the measurement of  $R_{K^*} \equiv \Gamma (B^0 \to K^{*0} \mu^+\mu^-)/\Gamma(B^0 \to K^{*0} e^+ e^-)$ \cite{rkstar}:
\bea
R_{K^*}^{[0.045,1.1]}& = &0.660^{+0.110}_{-0.070}~{\rm (stat)} \pm 0.024~{\rm (syst)}\,,\,\,\\
R_{K^*}^{[1.1,6.0]} &= &0.685^{+0.113}_{-0.069}~{\rm (stat)} \pm 0.047~{\rm (syst)},
\eea
where the superscript denotes the dilepton invariant mass-squared $q^2$ range.
These measurements differ from the SM prediction, which is  $\simeq 1$ \cite{Hiller:2003js, Bordone:2016gaq},
 by 2.2-2.4$\sigma$ in the low-$q^2$ region and  by
2.4-2.5$\sigma$ in the central-$q^2$ region.  In Moriond 2019, the Belle collaboration has presented their first measurements of $R_{K^*}$ in $B^0$ decays along with the first ever measurement of  $R_{K^*}$ in $B^+$ decays \cite{rkstar2019}. However, these measurements have large uncertainties due to which the $R_{K^*}$ disagreement with the SM is still at the level of  $2.4\sigma$.
Apart from $R_{K^{(*)}}$, there are other measurements, all in the $b \to s \, \mu^+ \, \mu^-$ sector, which show discrepancies with the SM. The measurement of some of the angular observables \cite{Kstarlhcb1,Kstarlhcb2,KstarBelle}, in particular $P'_5$, disagrees with the SM predictions \cite{sm-angular} at the level of 4$\sigma$ in the (4.3-8.68) $q^2$-bin. This disagreement is further supported by the measurements of ATLAS \cite{kstaratlas} and CMS \cite{kstarcms} collaborations. Also, there is tension in the branching ratio of $B_s \to \phi \mu^+ \mu^-$\cite{bsphilhc1,bsphilhc2}. Therefore  $b \to s \, \mu^+ \, \mu^-$  and $b \to c\, \tau\, \bar{\nu}$ decays  serve as a fruitful hunting ground to probe beyond SM physics.

In order to identify the Lorentz structure of new physics responsible for various anomalies in the $b \to s \, \mu^+ \, \mu^-$ sector, in a model-independent way, there have been a plethora of works in recent times \cite{rkstar-refs-before,rkstar-refs-after}.  For $b \to c\, \tau\, \bar{\nu}$, refs.  \cite{Freytsis:2015qca,Alok:2017qsi,Blanke:2018yud,Alok:2019uqc,Murgui:2019czp} identified new physics operators which can account for $R_{D^{(*)}}$ anomaly. However, simultaneous explanation of anomalies in  $b \to s \, \mu^+ \, \mu^-$  and $b \to c\, \tau\, \bar{\nu}$ sector in specific new physics models is bit tricky. This is because the $b \to c\, \tau\, \bar{\nu}$ transition occurs at the tree level within the SM, whereas $b \to s\, \mu^+\, \mu^-$ decay can only occur at the loop level. One needs a relatively large new physics contributions in order to explain the  $R_{D^{*}}$ anomaly. However such a large new physics contributions must also be consistent with the measurement of other observables which are in agreement with their SM predictions. Therefore there are only a limited set of new physics models which can simultaneously explain the  $R_{D^{(*)}}$ and $b \to s\, \mu^+\, \mu^-$ anomalies, see e.g. \cite{Bhattacharya:2014wla,Alonso:2015sja,Greljo:2015mma,Calibbi:2015kma,Fajfer:2015ycq,Bauer:2015knc,Barbieri:2015yvd,Boucenna:2016wpr,Das:2016vkr,Boucenna:2016qad,Deshpand:2016cpw,Becirevic:2016oho,Sahoo:2016pet,Bhattacharya:2016mcc,Crivellin:2017zlb,Chen:2017hir,Buttazzo:2017ixm,Megias:2017ove}.

In this work we revisit four models: (1) SM like vector bosons (VB), (2) $SU(2)_L$-singlet vector leptoquark ($U_1$), (3) $SU(2)_L$-triplet scalar leptoquark ($S_3$) and (4) $SU(2)_L$-triplet vector leptoquark  models ($U_3$). We assume a coupling to only third generation in the gauge basis.  These models were studied in \cite{Calibbi:2015kma,Bhattacharya:2016mcc}. In ref. \cite{Bhattacharya:2016mcc}, it was shown that VB model is a viable model, but the $Br(\tau \to 3\mu)$ were shown to be  $O(10^{-8})$ which is close to its present upper bound of $2.1 \times 10^{-8}$ \cite{Hayasaka:2010np}. Further $U_1$ model was also considered to be a potential model in refs. \cite{Calibbi:2015kma,Bhattacharya:2016mcc} to explain $R_K$ and $R_{D^{*}}$ anomalies.  In refs.~\cite{Feruglio:2016gvd,Feruglio:2017rjo}, it was shown that the $U_1$ model is disfavored due to the constraints coming from processes such as $\tau \to e \nu \bar{\nu}$, $\tau \to \mu \nu \bar{\nu}$ which are generated solely due to renormalization group running (RGE) effects. However, it should be noted that refs.~\cite{Feruglio:2016gvd,Feruglio:2017rjo} considered a different kind of transformation from gauge to mass basis as used in this work. We choose the one used in the ref. \cite{Bhattacharya:2016mcc}\footnote{Note that in the limit of very small mixing angles these two transformations are same.}. Further, in ref. \cite{Faroughy:2016osc}, it was shown that the high-$p_T$ searches also disfavor the $U_1$ model.

For the first time we perform a global analyses of the four NP models by taking into account  all relevant data from B-sector as well as the leptonic decays generated through the RGE running. Our main findings are summarized below:
\begin{enumerate}
\item  In ref. \cite{Bhattacharya:2016mcc}, it was found that the $VB$ is a viable model for the combined explanation of the B-anomalies. However, on the basis of 
global fit, we found that for the $VB$ model the $\mathcal{B}(\tau \to 3\mu)$ is an order magnitude above the current experimental upper bound at the best fit. Therefore, we show that the $VB$ model is simply excluded by $\tau \to 3\mu $ process.
\item Ref. \cite{Calibbi:2015kma} motivated $U_1$ LQ model as potential candidate for the combined explanation of the B-anomalies. Following this, ref. \cite{Bhattacharya:2016mcc} showed it to be a viable model. However, our analyses shows that $R_{D^{*}}$ value is only marginally improved over its SM value.
\item In refs \cite{Feruglio:2016gvd,Feruglio:2017rjo} it were shown that this minimal framework for the simultaneous explanation of the anomalies is challenged by the $\tau$ lepton flavor violating decays and Z-pole observables which arise due to RGE running. However, we find that using a more flexible transformation matrices to go from gauge to mass basis,   the constraints coming from these processes can be avoided.
\end{enumerate}

The  paper  is  arranged  as  follows.    After  the  introduction, in Sec. II, we discuss the methodology used in our anlysis. In Sec. III, we describe the four new physics models in the framework of third generation coupling in the weak basis.  In Sec.  IV, we present our results.  Finally, we conclude in Sec. V.

\section{\bf Methodology} \label{sec:method}
In our analyses we take into account  following  constraints

\begin{enumerate}
\itemsep0em 
\item the measurement of $R_{D^{(*)}}$ \cite{average,rdrdstarnew},
\item the branching ratio of $B^0_s \to \mu^+ \mu^-$ \cite{Aaij:2013aka,CMS:2014xfa,Aaboud:2018mst},
\item the differential branching ratio of $B^{0} \to K^{*0} \mu^+ \mu^-$ and  $B^{+} \to K^{*+} \mu^+ \mu^-$ measured by LHCb \cite{Aaij:2014pli,Aaij:2016flj},
\item the CP-averaged differential angular distribution for $B^0 \to K^{*0} (\to K^+ \pi^-) \mu^+ \mu^-$  \cite{Kstarlhcb2},
\item the differential branching ratio of $B^{0}\to K^{0} \mu^+\mu^-$ and $B^{+}\to K^{+} \mu^+\mu^-$ measured by LHCb \cite{Aaij:2014pli} and CDF \cite{CDFupdate},
\item  the differential branching ratio of $B^0_s  \to \phi \mu^+ \mu^-$  by LHCb \cite{bsphilhc2} and CDF \cite{CDFupdate} and the angular observables measured by LHCb \cite{bsphilhc2}, 
\item the differential branching ratio of $B \to X_s \mu^+ \mu^-$  measured by BaBar \cite{Lees:2013nxa},
\item the recent data by ATLAS \cite{BK*mumuATLAS} and CMS \cite{BK*mumuCMS} for the angular observables in $B^{0} \to K^{*0}\mu^+ \mu^-$ decay,
\item the measurement of $R_{K^{(*)}}$  \cite{Aaij:2019wad,rkstar,rkstar2019},
\item mass difference $\Delta M_s$ in $B_s$-mixing \cite{hfag},
\item branching ratio of $B \to K^{(*)} \nu \bar \nu$ \cite{Grygier:2017tzo}, 
\item branching ratio of $\tau \to 3 \mu$ \cite{Hayasaka:2010np},
\item branching ratio of $B \to K  \mu \tau$ \cite{Aubert:2007rn},
\item branching ratio of $\tau \to \phi \mu$ \cite{Miyazaki:2011xe},
\item branching ratio of $\tau \to e \nu \bar \nu$ \cite{hfag}.
\end{enumerate}

Note that out of the four models, not all of these contribute to all the observables. For example, the leptoquark models do not contribute to the four fermion operator at the tree level, i.e. the processes like branching ratio of $\tau \to 3 \mu$, $\tau \to e \nu \bar \nu$ and $B^0_s$-$\bar{B^0_s}$ mixing do not occur at the tree level in these models. Additionally, $U_1$ model does not give any new physics contribution to the processes related to $b \to s \nu \bar{\nu}$ transition  at the tree level.

In order to check this viability of the new physics models, we perform three kinds of fit,
\begin{enumerate}
\item Fit 1 : Global fit
\item Fit 2 : Fit with excluding $b\to c \tau \bar \nu$ data
\item Fit 3 : Fit with only clean observables
\end{enumerate}

In fit 1, we perform the global fit by taking all the relevant data from 1 to 10.  Using the fit results, we check the consistency of the other observables from 11 to 15. We then remove the $R_{D^{(*)}}$ data from the fit 1 to perform second kind of fit. Fit 2 would enable us to know that how well these models can explain the anomalies in $b \to s $ sector and on the other hand would shed light on the what $b \to s \, \mu^+ \, \mu^-$ data imply in $b \to c\, \tau\, \bar{\nu}$ 
sector in the context of these models.
In the third kind of fit, we only consider clean observables such as $R_{K^{(*)}}$, $R_{D^{(*)}}$. Along with this the model specific constraints such as $B^0_s$-$\bar{B^0_s}$ mixing is also taken into account. This fit is performed to check to what extent our conclusions are dependent on hadronic uncertainties.

We do a $\chi^2$ fit using CERN minimization code {\tt MINUIT} \cite{James:1975dr}. The  $\chi^2$  function is defined as
\beq
\chi^2(C_i) = (\mathcal{O}_{th}(C_i) -\mathcal{O}_{exp})^T \, \mathcal{C}^{-1} \,
(\mathcal{O}_{th}(C_i) -\mathcal{O}_{exp})\,.
\eeq  
The theoretical predictions, $\mathcal{O}_{th}(C_i)$ are calculated using {\tt flavio} \cite{Straub:2018kue}. The $\mathcal{O}_{exp}$ are the experimental measurements of the observables used in the fit.  The total covariance matrix $\mathcal{C}$ is obtained by adding the individual theoretical and experimental covariance matrices. We closely follow the methodology for global fits discussed in Refs. \cite{Alok:2017sui,Alok:2017jgr}.  

\section{\bf New physics models}
In this section we describe the general framework and discuss the new physics models which we have studied. In order to explain the $R_K$ anomaly, ref. \cite{Glashow:2014iga} considered following operator
\beq
\frac{G}{\Lambda^2}(\bar{b'_L}\gamma_{\mu}b'_L)(\bar{\tau'_L}\gamma^{\mu}\tau'_L).
\label{GGL}
\eeq
Here $\Lambda$ is the scale of new physics. It was assumed that, in the gauge basis, NP couple to the third generation only. The transformation from gauge basis to the mass basis will then generate operator 
$(\bar{s}_L\gamma_{\mu}b_L)(\bar{\mu}_L\gamma^{\mu}\mu_L)$ which contributes to $b \to s \mu^+ \mu^-$.

Then in ref. \cite{Bhattacharya:2014wla} it was pointed out that full $SU(3)_C \times SU(2)_L \times U(1)_Y$ invariant operator can be written in the following way
\begin{equation}
{\cal L}_{eff} = \frac{G_1^{ijkl}}{\Lambda^2}(\bar{Q'^{i}_L}\gamma_{\mu}Q'^{j}_L)(\bar{L'^{k}_L }\gamma^{\mu}L'^{l}_L)
 + \frac{G_2^{ijkl}}{\Lambda^2}(\bar{Q'^{i}_L}\gamma_{\mu}\sigma^I Q'^{j}_L)(\bar{L'^{k}_L}\gamma^{\mu}\sigma^I L'^{l}_L),
 \label{lnp}
\end{equation}
the $i, j, k$ and $l$ are the generation indices and $Q'$ and $L'$ are 
the quark and lepton doublets in the gauge basis. The four fermion operator in the first term of Eq. \ref{lnp} contains neutral current interactions only while the second four fermion operator contains both charged and neutral current interactions and hence can simultaneously generate NP effects in both $R_{K^{(*)}}$ and $R_{D^{(*)}}$.

 We assume that the only non-zero Wilson coefficients are $G_1^{3333}$ and $G_2^{3333}$, ensuring that
the NP couples to only the third generation in the gauge basis. The transformation used in going from gauge basis to mass basis is given by
\begin{equation} 
u'_L = Uu_L ,~~ ~d'_L = D d_L,~~~ l'_L = L l_L, ~~~\nu'_L = L\nu_L, 
\end{equation}
here the primed spinors (gauge basis) has all three generation 
of fermions and $U$, $D$, and $L$ are 3$\times$3 unitary matrices. 
Note that the transformation for both the charged and neutral leptons 
are assumed to be the same because the neutrino masses are neglected here. 

Based on the effective operators, the global analyses indicate that a non-zero new physics contribution to operator $(\bar c_L \gamma^\mu b)(\bar \nu_L \gamma^\mu \tau_L)$ is required to explain the $R_{D^{(*)}}$ anomalies. On the other hand, for explaining the anomalies in $R_{K^{(*)}}$ and $b\to s \mu^+ \mu^-$ data, we need a new physics in $(\bar s_L \gamma^\mu b)( \bar \mu \gamma_\mu \mu)$ operator. Since both the operators involve only second and third generations, we assume mixing only between second and third 
generation  so that the matrices $D$ and $L$ can  be defined using the two rotation angles  $\theta_{bs}$ and $\theta_{\mu \tau}$, respectively. 

Therefore, we define
$$
 D=
  \left( {\begin{array}{ccc}
   1 & 0 & 0\\
   0 & \cos{\theta_{bs}} &\sin {\theta_{bs}}\\
   0 & -\sin{\theta_{bs}}   & \cos{\theta_{bs}}\\
  \end{array} } \right), \,\,
 L=
  \left( {\begin{array}{ccc}
   1 & 0 & 0\\
   0 & \cos{\theta_{\mu \tau}} &\sin {\theta_{\mu \tau}}\\
   0 & -\sin{\theta_{\mu \tau}}   & \cos{\theta_{\mu \tau}}\\
  \end{array} } \right)$$.\\

The transition $b \to c \tau \bar \nu$ occurs at the tree level in the SM so the explanation of anomalies in $R_{D^{(*)}}$ requires a large NP coupling with third generation. However, as the transition $b \to s \mu \mu$ occurs at loop level in SM,  a small NP coupling with second generation is required for the explanation of $R_{K^{(*)}}$ anomalies. The above  transformation gives small couplings through rotation to the second generation down quarks and charged leptons in the mass basis. 
 
Therefore,  in the mass basis, 
the new physics couplings can be written as,
\begin{equation}
\centering
G_{(1,2)}^{ijkl} = g_{(1,2)}X^{ij}Y^{kl},
\end{equation}
where $X$ and $Y$ are the matrices which are function of the rotation angles. The form of these matrices for 
$b \to s l^+ l^-$ decay is 
\begin{equation}
   X= D^\dagger \left[ {\begin{array}{ccc}
   0 & 0 & 0\\
   0 & 0 & 0\\
   0 & 0 & 1 \\
  \end{array} } \right]D = 
  \left[ {\begin{array}{ccc}
   0 & 0 & 0\\
   0 & \sin^2{\theta_{bs}} &- \sin {\theta_{bs}} \cos {\theta_{bs}}\\
   0 & -\sin{\theta_{bs}} \cos {\theta_{bs}}  & \cos^2{\theta_{bs}}\\
  \end{array} } \right],
\end{equation}
\begin{equation}
   Y= L^\dagger \left[ {\begin{array}{ccc}
   0 & 0 & 0\\
   0 & 0 & 0\\
   0 & 0 & 1 \\
  \end{array} } \right]L = 
  \left[ {\begin{array}{ccc}
   0 & 0 & 0\\
   0 & \sin^2{\theta_{\mu \tau}} & -\sin {\theta_{\mu \tau}} \cos {\theta_{\mu \tau}}\\
   0 & -\sin{\theta_{\mu \tau}} \cos {\theta_{\mu \tau}} & \cos^2{\theta_{\mu \tau}}\\
  \end{array} } \right].
\end{equation}
In case of up-type quarks involved in the process, the matrix $U$ is used instead of $D$.

The couplings $g_1$ and $g_2$ take specific values depending upon the new physics models. 
In the more recent study \cite{Kumar:2018kmr} generic couplings in the mass basis 
are considered.

The effective Hamiltonian relevant for the 
$b\to s\ell^+_i\ell^-_j $, $b \to c \ell_i \bar{\nu}_j$ and $b \to s \nu_i \bar \nu_j$ processes can be written as
\bea
H_{eff}(b\to s\ell^+_i\ell^-_j ) & =&  - \frac{\alpha G_F}{\sqrt{2}\pi}V_{tb}V_{ts}^*\Big[C_9^{ij}(\bar{s}_L\gamma^{\mu}b_L)(\bar{l_i}\gamma_{\mu}l_j)+~C_{10}^{ij}(\bar{s}_L\gamma^{\mu}b_L)(\bar{l_i}\gamma_{\mu}\gamma^5 l_j)\Big],\\
H_{eff}(b \to c \ell_i \bar{\nu}_j) &=& \frac{4G_F}{\sqrt{2}}V_{cb}C_{V}^{ij}(\bar{c}_L\gamma^{\mu}b_L)(\bar{l}_{iL}\gamma_{\mu}\nu_{jL}), \\
H_{eff}(b \to s \nu_i \bar \nu_j)&=&-\frac{\alpha G_F}{\sqrt {2} \pi} V_{tb} V_{ts}^*
C_L^{ij} (\bar s_L \gamma^\mu b_L) (\bar \nu_i \gamma_\mu (1-\gamma^5) \nu_j).
\eea
The new physics contributions to the Wilson coefficients read
{
\bea
C_9^{\mu\mu} &=& -C_{10}^{\mu\mu} =- \frac{\pi}{\sqrt{2}\alpha G_F V_{tb}V_{ts}^*}\frac{(g_1 + g_2)}{\Lambda^2} \left(\sin\theta_{bs}\cos\theta_{bs}\sin^2\theta_{\mu\tau}\right),
\label{c9mumu}\\
C_V^{ij} &=& -\frac{1}{2\sqrt{2}G_F V_{cb}}\frac{2g_2}{\Lambda^2}
\Big(-V_{cs}\sin\theta_{bs}\cos\theta_{bs}  + V_{cb} \cos^2\theta_{bs}\Big)Y^{ij},  \\
C_L^{ij}& = & -\frac{\pi}{ \sqrt 2 \alpha G_F V_{tb} V_{ts}^*} \frac{(g_1-g_2)}{\Lambda^2}\left(\sin \theta_{bs} \cos \theta_{bs} \right) Y^{ij} .
\label{wcbsnunu}
\eea
}
Having discussed our general framework and assumptions, next we discuss the four new physics models studied in this paper.
\subsection{SM-like vector bosons (VB)} 
\label{VBmodel}
\smallskip
First, we consider an additional heavy vector bosons which transforms as 
$({\bf1},{\bf 3},{\bf 1})$ under the SM gauge group  
$SU(3)_C \times SU(2)_L \times U(1)_Y$. In the gauge basis, its
interaction with the fermions is given by
\bea
\Delta{\cal L}^{}_{V} = g^{33}_{qV}\ ({\overline Q'_{L3}} \gamma^{\mu}\sigma^I Q'{_{L3}})V_{\mu}^I + g^{33}_{lV}\ ({\overline L'_{L3}}\gamma^{\mu}\sigma^I L^{'}_{L3})V_{\mu}^I.
\eea
On integrating out this heavy vector boson, this generates one of the two operators 
shown in Eq. \ref{lnp} with
\beq
g_1 = 0,~~~~~~ g_2 = - g^{33}_{qV} g^{33}_{\ell V}.
\eeq
For simplicity, we set these couplings to a fixed 
value as $g_{qV}^{33} = g_{lV}^{33} = \sqrt{0.5}$.
The $Z'$ (i.e the neutral component) contributes $b \to s \mu^+ \mu^-$ and $b \to s \nu {\bar\nu}$ decays, whereas the $W'$(the charged component) contributes to $b \to c \tau^- {\bar\nu}$ decay the at the tree level. In addition to the semileptonic operators, this model also generates four fermion operators at the tree level. These put additional constraints on the VB model.
For example, the $\Delta F=2$ process, the $B_s- \bar{B_s}$ mixing can be described by the 
effective Hamiltonian,
\bea
H_{eff}(B_s-\overline B_s) = \frac{G_F^2 m_W^2}{16\pi^2}(V_{tb} V_{ts}^*)^2 C_{VLL} (\bar s_L \gamma^\mu b_L)  (\bar s_L \gamma^\mu b_L).
\eea 
Here the Wilson coefficient is given by
\bea
C_{VLL} = C_{VLL}^{SM} + \frac{(g_{qV}^{33})^2}{2m_V^2}\frac{16\pi^2}{G_F^2 m_W^2(V_{tb}V_{ts}^*)^2} \sin^2\theta_{bs}\cos^2\theta_{bs}. \nn
\eea
the SM contribution to the Wilson coefficient $C_{VLL}^{SM}$ reads,
\beq
C_{VLL}^{SM} = \eta_{B_s}x_t\Big[1+\frac{9}{1-x_t} - \frac{6}{(1-x_t)^2} -\frac{6x_t^2\log{x_t}}{(1-x_t)^3}\Big].
\eeq
The QCD constant $\eta_{B_s}$ is equal to 0.551 and the ratio $x_t =m_t^2/m_W^2$.
The mass difference is given by
\bea
\Delta M_s = \frac{2}{3} ~m_{B_s} f_{B_s}^2 \hat B_{B_s}~ \frac{G_F^2 m_W^2}{16\pi^2}(V_{tb} V_{ts}^*)^2  C_{VLL}
\eea
As mentioned above, the VB also gives tree level contribution to the
 four lepton operators such as $(\bar \mu_L \gamma^\mu \tau_L) 
(\bar \mu_L \gamma_\mu \mu_L)$, which induces LFV tau-decay $\tau \to 3 \mu$.
The  branching ratio of $\tau \to 3 \mu$ is
\bea
B(\tau \to 3 \mu) = 0.94 \frac{0.5^2}{16 m_V^4} \frac{m_\tau^5 \tau_\tau}{192 \pi^3}  \sin^6\theta_{\mu\tau} \cos^2 \theta_{\mu\tau}.
\eea
On the experimental side, there is an upper bound on the branching ratio
$B(\tau \to 3 \mu)$, which is  $2.8 \times 10^{-8}$ at $90\%$ 
CL \cite{Hayasaka:2010np}.  In addition the VB model can contribute to more processes such as $\tau \to \mu \nu \bar \nu$, $\tau \to \mu \rho$ etc. But, we will show that the model is excluded if we take into account just the constraints from $R_{K^{(*)}}$, $R_{D^{(*)}}$, $\Delta M_s$ and $Br(\tau \to 3\mu)$.

\subsection{Leptoquark(LQ) models}
We consider the three leptoquark models, a scalar $SU(2)_L$ singlet LQ $S_1$ $(3,1,-2/3)$, a scalar triplet LQ $S_3$$(3,3,-2/3)$ and vector singlet LQ $U_1$$(3,1,4/3)$. In the gauge basis, the interaction Lagrangian for these LQ models is given by \cite{Sakaki:2013bfa}
\bea
\Delta{\cal L}^{}_{U_1} &=& g_{U_1}^{33}({\overline Q'}_{L3}\gamma^{\mu}L'_{L3})U_{1\mu} +{\rm h.c.} ~,\\ 
\Delta{\cal L}^{}_{S_3} &=& g_{S_3}^{33}( {\overline Q'_{L3}}
\sigma^I i \sigma^2  L_{L3}^{'c}) S_{3}^I +{\rm h.c.} ~,\\
\Delta{\cal L}^{}_{U_3} &=& g_{U_3}^{33}({\overline Q'}_{L3}\gamma^{\mu} \sigma^I L'_{L3})U_{3\mu} +{\rm h.c.}~.
\eea
On integrating out a heavy LQ, the operators of Eq. \ref{lnp} are generated at the tree level but with different weights of the two operators depending on the representation. We can identify the couplings $g_1$ and $g_2$ for various LQ models as
\bea
U_1:~ g_1 &=& g_2 = - \frac12  |g_{U_1}^{33} |^2 < 0~, \\
S_3: g_1 &=& 3g_2 =  \frac34  |g_{S_3}^{33} |^2 > 0~, \\
U_3: g_1 &=&-3g_2 = - \frac32  |g_{U_3}^{33} |^2 < 0,
\eea
the couplings $g_{U_1}^{33}$, $g_{U_3}^{33}$ and $g_{S_3}^{33}$ are set to one.
Clearly, all these LQ models can potentially contribute to  
$b \to s \mu^+ \mu^-$, $b \to s \nu {\bar\nu}$ and $b \to c \tau^- {\bar\nu}$
transitions at the tree level. However, for the $U_1$ model since $g_1=g_2$, there is no tree level NP contribution to $b \to s \nu {\bar\nu}$ (see Eq. \ref{wcbsnunu}). Note that the LQ models do not give contributions to the four fermion operators at the tree level. But these operators are still generated at low scale due to RGE running of the operators of Eq. \ref{lnp}~ \cite{Feruglio:2016gvd,Feruglio:2017rjo}. Some of the processes which are generated due to RGE effects in the LQ models are $B \to K^{(*)} \nu {\bar\nu}$, 
$\tau \to 3 \mu$, $\tau \to \mu \nu \bar \nu$, $\tau \to e \nu \bar \nu$, $\tau \to \mu \rho$.
In addition to this the Z-boson axial and vector couplings are also affected. It was shown in Refs. ~\cite{Feruglio:2016gvd,Feruglio:2017rjo} that the $Br(\tau \to e \bar\nu \nu)$ 
constraints can be very stringent for the models with $g_1 = g_2$. However, note that in the present work we use a different transformation to rotate from gauge to mass basis. 

 \begin{table*}[!h]
\centering
  \resizebox{\textwidth}{!}{  
\begin{tabular}{|c|c|c|c|c|c|c|c|c|c|c|} \hline
\multicolumn{2}{|c|}{Observable } & ${R_K}_{[1.1-6.0]}$ 
& ${R_{K^*}}_{[0.045-1.1]}$ & ${R_{K^*}}_{[1.1-6.0]}$& ${P'_{5}}_{[4.0-6.0]}$ & $R_D^{ratio}$ & $R_{D^*}^{ratio}$ &$\frac{B_K^{SM+NP}}{B_K^{SM}}$ & $\frac{B_{K^*}^{SM+NP}}{B_{K^*}^{SM}}$&$\tau \to 3 \mu $  \\ \hline
\multicolumn{2}{|c|}{Measurement }  &$0.846 \pm 0.062$ 
&$0.66 \pm 0.09$ &$0.69 \pm 0.10$ & $-0.30 \pm 0.16 $ & $1.10 \pm 0.10$ &$1.16 \pm 0.06$ &$\leq 4.0$&$\leq 2.9$&$<2.1\times 10^{-8}$   \\ \hline
\multicolumn{2}{|c|}{Standard Model }  &1.0 & 0.93 & 0.99 &-0.82& 1.0 & 1.0& &&  \\ \hline
$\theta_{\mu\tau}(rad.)$ & $\theta_{bs}(rad.)$   & \multicolumn{8}{|c|}{Fit 1(Global fit)}&    \\ \hline
$ 0.649 \pm 0.158$&$ 0.004 \pm 0.002$& 0.78  & 0.87 & 0.79 &-0.71&1.02  &1.02&0.88  &0.88 &$5.95\times 10^{-7}$ \\ \hline
\multicolumn{2}{|c|}{$\chi^2$} & 1.09  &5.11 &0.99&5.20 & 0.54  &4.56&-- &-- &\\ \hline
$\theta_{\mu\tau}(rad.)$ & $\theta_{bs}(rad.)$   & \multicolumn{8}{|c|}{Fit 2}&    \\ \hline
$ 1.53 \pm 2.47$&$ 0.002 \pm 0.004$& 0.71  & 0.85 & 0.72 &-0.68&0.93  &0.93&0.94  &0.94 &$3.18\times 10^{-8}$\\ \hline
\multicolumn{2}{|c|}{$\chi^2$} & 4.98  &4.27 &0.10&4.61 & 2.53  &13.64&-- &-- & \\ \hline
$\theta_{\mu\tau}(rad.)$ & $\theta_{bs}(rad.)$   & \multicolumn{8}{|c|}{Fit 3}&  \\ \hline
$ 0.535 \pm 0.16$&$ 0.005 \pm 0.003$& 0.80  & 0.88 & 0.81 &-0.71&1.03  &1.03&0.85  &0.84 &$2.5\times 10^{-7}$\\ \hline
\multicolumn{2}{|c|}{$\chi^2$} & 0.45  &5.38 &1.47&5.20 & 0.43  &3.99&-- &-- & \\ \hline
\end{tabular}
}
\caption{Best fit values of the mixing angles in vector boson model. For Fit 1, dof=123,  $\chi^2/dof = 1.00$.} 
\label{tab:results_vb_all}
\end{table*}


\begin{table}[!h]
\centering
\resizebox{\textwidth}{!}{ 
\begin{tabular}{|c|c|c|c|c|c|c|c|} \hline
\multicolumn{2}{|c|}{Observable } & ${R_K}_{[1.1-6.0]}$ 
& ${R_{K^*}}_{[0.045-1.1]}$ & ${R_{K^*}}_{[1.1-6.0]}$ & ${P'_{5}}_{[4.0-6.0]}$& $R_D^{ratio}$ & $R_{D^*}^{ratio}$   \\ \hline
\multicolumn{2}{|c|}{Measurement }  &$0.846 \pm 0.062$ 
&$0.66 \pm 0.09$ &$0.69 \pm 0.10$ & $-0.30 \pm 0.16 $ &  $1.10 \pm 0.10$ &$1.16 \pm 0.06$ \\ \hline
\multicolumn{2}{|c|}{Standard Model }  &1.0 & 0.93 & 0.99 & -0.82 & 1.0 & 1.0   \\ \hline
$\theta_{\mu\tau}(rad.)$ & $\theta_{bs}(rad.)$   & \multicolumn{6}{|c|}{Fit 1(Global fit)}   \\ \hline
$ 0.339 \pm 0.224$&$ 0.007 \pm 0.009$& 0.77  & 0.88 & 0.78 &-0.70&1.04  &1.04 \\ \hline
\multicolumn{2}{|c|}{$\chi^2$} & 1.54  &5.38 &0.79&5.11 & 0.30  &3.71\\ \hline
$\theta_{\mu\tau}(rad.)$ & $\theta_{bs}(rad.)$   & \multicolumn{6}{|c|}{Fit 2}   \\ \hline
$ 0.095 \pm 0.134$&$ 0.0895 \pm 0.255$& 0.76  & 0.87 & 0.77 &-0.70&0.94  &0.94 \\ \hline
\multicolumn{2}{|c|}{$\chi^2$} & 1.80  &5.11 &0.69&5.11 & 2.45  &13.35\\ \hline
$\theta_{\mu\tau}(rad.)$ & $\theta_{bs}(rad.)$   & \multicolumn{6}{|c|}{Fit 3}  \\ \hline
$ 0.328 \pm 0.21$&$ 0.007 \pm 0.008$& 0.78  & 0.87 & 0.79 &-0.71&1.04  &1.04 \\ \hline
\multicolumn{2}{|c|}{$\chi^2$} & 1.09  &5.11 &0.99&5.20 & 0.30  &3.71\\ \hline
\end{tabular}
}
\caption{Best fit values of the mixing angles in $U_1$ leptoquark model. For Fit 1, dof=122,  $\chi^2/dof = 1.00$.}
\label{tab:results_u1_all}
\end{table}
\begin{table}[!h]
\centering
  \resizebox{\textwidth}{!}{ 
\begin{tabular}{|c|c|c|c|c|c|c|c|c|c|} \hline
\multicolumn{2}{|c|}{Observable } & ${R_K}_{[1.1-6.0]}$ 
& ${R_{K^*}}_{[0.045-1.1]}$ & ${R_{K^*}}_{[1.1-6.0]}$& ${P'_{5}}_{[4.0-6.0]}$ & $R_D^{ratio}$ & $R_{D^*}^{ratio}$ &$\frac{B_K^{SM+NP}}{B_K^{SM}}$ & $\frac{B_{K^*}^{SM+NP}}{B_{K^*}^{SM}}$  \\ \hline
\multicolumn{2}{|c|}{Measurement }  &$0.846 \pm 0.062$  
&$0.66 \pm 0.09$ &$0.69 \pm 0.10$ & $-0.30 \pm 0.16 $ &  $1.10 \pm 0.10$ &$1.16 \pm 0.06$ &$\leq 4.0$&$\leq 2.9$   \\ \hline
\multicolumn{2}{|c|}{Standard Model }  &1.0 & 0.93 & 0.99 &-0.82& 1.0 & 1.0& &  \\ \hline
$\theta_{\mu\tau}(rad.)$ & $\theta_{bs}(rad.)$   & \multicolumn{8}{|c|}{Fit 1(Global fit)}   \\ \hline
$1.57 \pm 0.48$  &$-0.0008 \pm 0.0001 $ & 0.76  & 0.87 & 0.77 &-0.70 &1.04&1.04&1.02  &1.02\\ \hline
\multicolumn{2}{|c|}{$\chi^2$} &1.80 &5.11 &0.69 & 5.11 & 0.30&3.71&-- &--  \\ \hline
$\theta_{\mu\tau}(rad.)$ & $\theta_{bs}(rad.)$   & \multicolumn{8}{|c|}{Fit 2}   \\ \hline
$0.095 \pm 0.134$  &$-0.089 \pm 0.255 $ & 0.76  & 0.86 & 0.77&-0.70 &0.91&0.91&10.82  &10.82 \\ \hline
\multicolumn{2}{|c|}{$\chi^2$} &1.80 &4.84 &0.69 & 5.11 & 3.32&16.75&-- &--  \\ \hline
$\theta_{\mu\tau}(rad.)$ & $\theta_{bs}(rad.)$   & \multicolumn{8}{|c|}{Fit 3}   \\ \hline
$1.57 \pm 0.48$  &$-0.0007 \pm 0.0002 $ & 0.79  & 0.87 & 0.80&-0.71 &1.04&1.04&1.02  &1.02 \\ \hline
\multicolumn{2}{|c|}{$\chi^2$} &0.82 &5.11 &1.16 & 5.20 & 0.30&3.71&-- &--  \\ \hline
\end{tabular}
}
\caption{Best fit values of the mixing angles in $S_3$ leptoquark model.For Fit 1, dof=122,  $\chi^2/dof = 1.02$.}
\label{tab:results_s3_all}
\end{table}
\section{\bf Results and discussions}
In this section, we present the result of the fits for the SM like vector boson, and $U_1$, $S_3$, $U_3$ LQ models. Their mass set to 1 TeV. 
As discussed in Sec.~\ref{sec:method}, we perform three kind of fits. For the first one, which we call a Global fit, all relevant data is included. 
We then perform a fit, fit 2,  by removing only the $R_{D^{(*)}}$ data from the $\chi^2$.  Finally, a fit including only the clean observables is performed.
The fit results for the VB model are presented in Table \ref{tab:results_vb_all}, and we observe the following:
At the best fit point the VB model evades the current upper bound on the $Br(\tau \to 3\mu)$ and this holds even on removing the $b \to c \tau \bar{\nu}$ data from the fit. Therefore, we can conclude that the VB model, coupling only to the third generation in the weak basis is inconsistent with the present data.

\begin{figure}[htbp]
\centering
\resizebox{\textwidth}{!}{ 
\begin{tabular}{cc}
\includegraphics[]{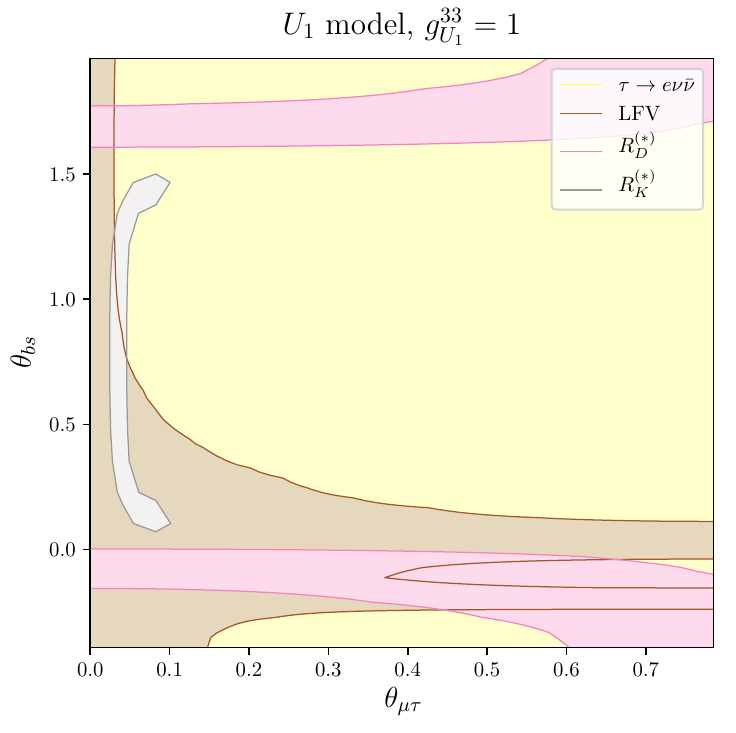}& \includegraphics[]{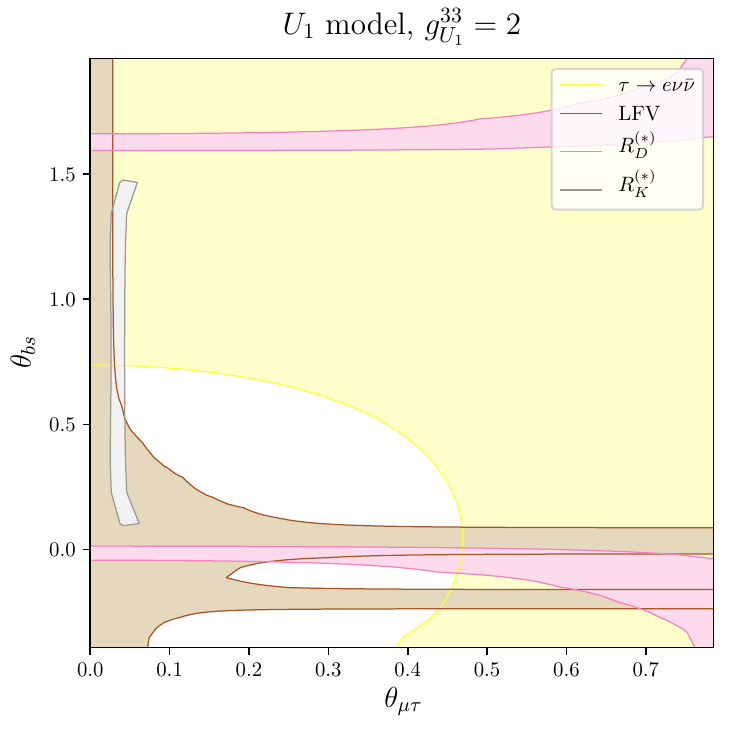}  \\
\includegraphics[]{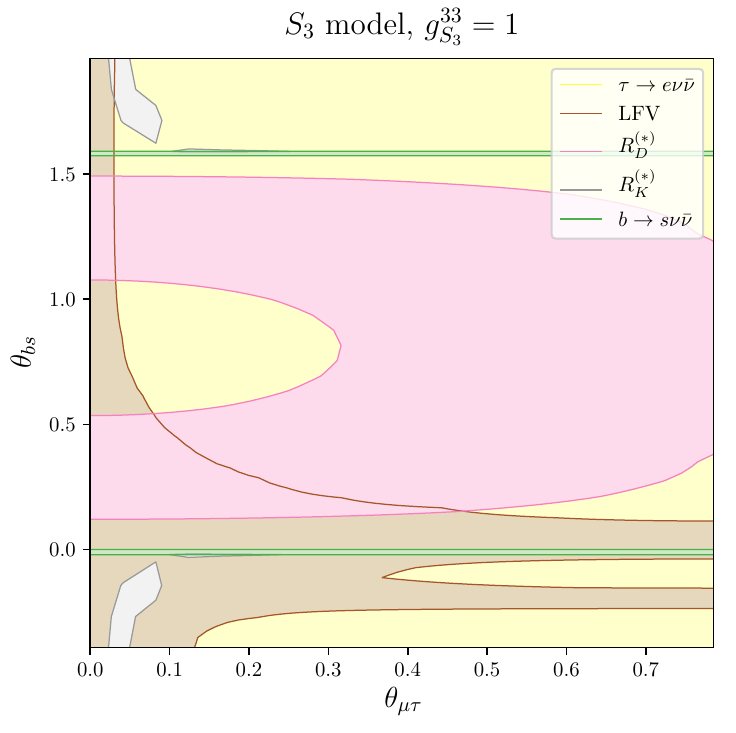} &\includegraphics[]{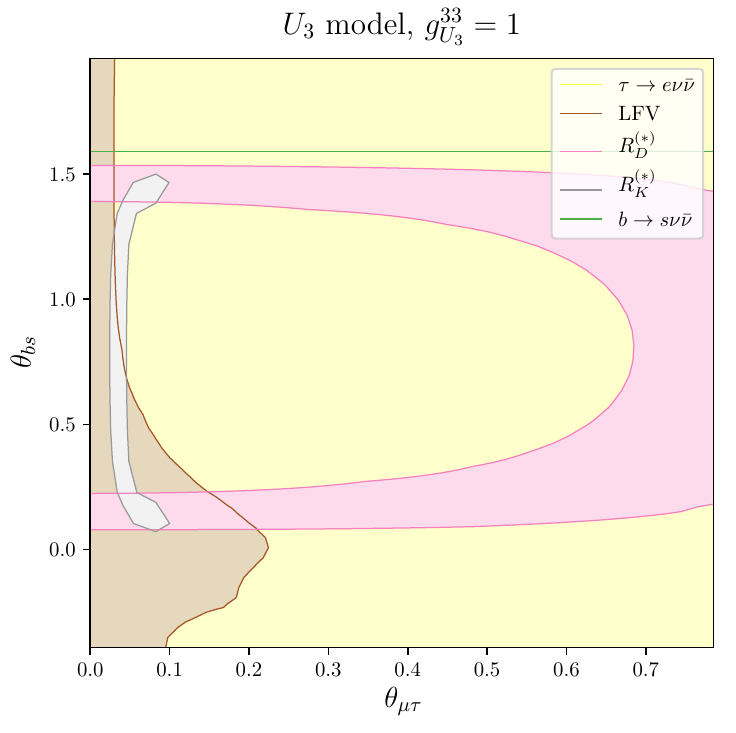}\\
\end{tabular}
}
\caption{(Color Online) The  plots  in the top row depict allowed new physics parameter space, at 2$\sigma$, due to various observables in the $U_1$ model. The left and right panels in the top row represent the allowed regions for $g_{U_1}^{33} = 1$ and $g_{U_1}^{33} = 2$ , respectively.  The left and right panels in the bottom row depicts the allowed regions in the $S_3$ and $U_3$ models, respectively. The plots for $S_3$ and $U_3$ models correspond to  $g_{S_3(U_3)}^{33} = 1$. The grey, magenta, brown, yellow and green regions in these figures show allowed parameter space for $R_{K^{(*)}}$, $R_{D^{(*)}}$, branching ratios of LFV decays ($\tau \to \phi \mu$ \& $B\to K\tau^{\pm}\mu^{\mp}$), branching ratio of $\tau \to e \nu \bar \nu$ and branching ratios of $B \to K^{(*)} \nu \bar \nu$  transitions, respectively. }
\label{fig1}
\end{figure}

The fit results for the $U_1$ model are presented in Table \ref{tab:results_u1_all}.  At the best point the central values of $R_{K^{(*)}}$ in the central 
$q^2$ region lies within 1$\sigma$ of the experimental range, whereas this is not true for the $R_{K^{*}}$ in low $q^2$ bin. The central value of $R_{D^{*}}$ falls in 2$\sigma$ of the experimental range. Further, the angular observable $P'_5$ can be accommodated within 2$\sigma$. We get the similar results for the fits performed only with the clean observables.  On removing  $R_{D^{(*)}}$ from the fit, we find that the tension in the same gets even worse as compared to the SM. 

The new physics coupling are found to be highly correlated with large errors. Hence, one should also consider  the allowed range of various observables. In Fig.~ \ref{fig1} we present the $2\sigma$ contours of various observables in the space of couplings.

It is evident from the left panel of top row of Fig.~\ref{fig1} that the $R_{K^{(*)}}$ (grey) and $R_{D^{(*)}}$ (magenta) regions do not overlap even at $2\sigma$. The brown region depicts constraints on this model due to LFV decays, $Br(\tau \to \phi\mu)$ and $Br(B\to K\tau^{\pm}\mu^{\mp})$. Evidently, these puts a tight constraints on this model. On the other hand, the yellow region shows that the $Br(\tau \to e \nu \bar \nu)$ does not put additional constraint on this model. Note that for a  different mixing pattern, the Refs.~\cite{Feruglio:2016gvd,Feruglio:2017rjo} found this to be extremely constraining for the $U_1$ model which otherwise was found to be viable in their setup.

The main reason is that our parametrization is more flexible. In particular, we obtain the transformation matrices Eq. 18 of \cite{Feruglio:2016gvd,Feruglio:2017rjo} only in the limit $\theta_{bs}\to 0$ and $\theta_{\mu\tau} \to 0$. However, in our analysis we allow the parameters $\theta_{bs}$ and $\theta_{\mu \tau}$ to be free and fit them from the real data.  Our parametrization is more suitable because it provides more flexibility to satisfy the additional constraints such as $\mathcal{B}(\tau \to e \nu \bar \nu)$. To understand this, we note that according to Eq. 97 of Ref. \cite{Feruglio:2016gvd,Feruglio:2017rjo} this branching ratio is driven by the Wilson coefficient $(C_{L}^{\tau \ell})_{31} \propto g_2 \lambda^{u}_{33}  \lambda_{33}^e$ which according to \cite{Feruglio:2016gvd,Feruglio:2017rjo} is $ \simeq -0.5$, simply because they have set $\lambda^u_{33}=\lambda^e_{33}=1$ and $g_2=-0.5$ for U1 leptoquark. In contrast, according to our parametrization we get $(C_{L}^{\tau \ell})_{31} \sim g_2 \cos^2{\theta_{bs}} \cos^2{\theta_{\mu \tau}}$, which is a function of the mixing angles. Furthermore,  in our analyses the parameter $g_2$ is fixed by model but in a general EFT analysis \cite{Feruglio:2016gvd,Feruglio:2017rjo}, $C_3$ ($=g_2$) is a free parameter and  varying $C_3$ it as done in \cite{Feruglio:2016gvd,Feruglio:2017rjo}, is not equivalent to varying $\theta_{bs}$ and $\theta_{\mu \tau}$ because the later parameters also appear in all other observables such as $R_K$, $R_D$  with a different functional dependence.

The upper right panel in Fig.\ref{fig1} shows 
the same plot for $g_{U_1}^{33} = 2$, clearly the higher  values of the coupling also does not help. Therefore, we conclude that the $U_1$ LQ, in this present setup is not a viable model for the combined explanation of the charge and neutral current anomalies.

The fit results for $S_3$ and $U_3$ models are presented in the Tables
 \ref{tab:results_s3_all} and \ref{tab:results_u3_all} respectively. From these one can infer that: these models are able to explain  $R_{K^{(*)}}$ within 1$\sigma$, and reduces the tension in the $R_{D^{(*)}}$. The central value of $R_{K^{*}}$ for the low $q^2$  does not come within 1$\sigma$.
The angular observable $P'_5$ can be accommodated within $2 \sigma$. The tension in  $R_{D^{(*)}}$ becomes worse on removing $b \to c \tau \bar{\nu}$ data from the fit. In the fit with only clean observables,  the results of 
the global fit are almost unchanged.
The  left and right panels of in the bottom row of fig.~ \ref{fig1} represents the $2\sigma$ contours of the relevant observables in the couplings plane for $S_3$ and $U_3$ models, respectively. In $U_3$ model, the $R_{K^{(*)}}$ and $R_{D^{(*)}}$ regions do overlap within $2\sigma$, but this is challenged by the constraint coming from the upper bound on $b \to s \nu \bar \nu$ transitions. Again, we find that the constraints coming from the LFV decays are quite stringent. Further, we find that the  $Br(\tau \to e \nu \bar \nu)$ is not important for these models.

\begin{table}[t]
\centering
  \resizebox{\textwidth}{!}{ 
\begin{tabular}{|c|c|c|c|c|c|c|c|c|c|} \hline
\multicolumn{2}{|c|}{Observable } & ${R_K}_{[1.1-6.0]}$ 
& ${R_{K^*}}_{[0.045-1.1]}$ & ${R_{K^*}}_{[1.1-6.0]}$& ${P'_{5}}_{[4.0-6.0]}$ & $R_D^{ratio}$ & $R_{D^*}^{ratio}$ &$\frac{B_K^{SM+NP}}{B_K^{SM}}$ & $\frac{B_{K^*}^{SM+NP}}{B_{K^*}^{SM}}$   \\ \hline
\multicolumn{2}{|c|}{Measurement }  &$0.846 \pm 0.062$  
&$0.66 \pm 0.09$ &$0.69 \pm 0.10$ & $-0.30 \pm 0.16 $&  $1.10 \pm 0.10$ &$1.16 \pm 0.06$ &$\leq 4.0$&$\leq 2.9$   \\ \hline
\multicolumn{2}{|c|}{Standard Model }  &1.0 & 0.93 & 0.99&-0.82 & 1.0 & 1.0& &  \\ \hline
$\theta_{\mu\tau}(rad.)$ & $\theta_{bs}(rad.)$   & \multicolumn{8}{|c|}{Fit 1(Global fit)}   \\ \hline
$0.076 \pm 0.011$  &$0.144 \pm 0.036 $ & 0.76  & 0.87& 0.77&-0.70&1.14&1.14&296.27  &296.27 \\ \hline
\multicolumn{2}{|c|}{$\chi^2$} &1.80 &5.11 &0.69 &5.11& 0.18&0.06&-- &--  \\ \hline
$\theta_{\mu\tau}(rad.)$ & $\theta_{bs}(rad.)$   & \multicolumn{8}{|c|}{Fit 2}   \\ \hline
$0.083 \pm 0.117$  &$0.119 \pm 0.34 $ & 0.76  & 0.86 & 0.76&-0.70&1.11&1.11&207.14  &207.14 \\ \hline
\multicolumn{2}{|c|}{$\chi^2$} &1.80 &4.84 &0.79 &5.11& 0.006&0.72&-- &--  \\ \hline
$\theta_{\mu\tau}(rad.)$ & $\theta_{bs}(rad.)$   & \multicolumn{8}{|c|}{Fit 3}   \\ \hline
$0.071 \pm 0.013$  &$0.144 \pm 0.036 $ & 0.78  & 0.87 & 0.79&-0.71&1.14&1.14&296.27  &296.27 \\ \hline
\multicolumn{2}{|c|}{$\chi^2$} &1.09 &5.11 &0.99 &5.20& 0.18&0.06&-- &--  \\ \hline
\end{tabular}
}
\caption{Best fit values of the mixing angles  in $U_3$ leptoquark model, For Fit 1, dof=122,  $\chi^2/dof = 0.97$.}
\label{tab:results_u3_all}
\end{table}

\begin{table}[t]
\centering
  \resizebox{\textwidth}{!}{ 
\begin{tabular}{|c|c|c|c|c|c|c|c|c|c|c|c|} \hline
\multicolumn{4}{|c|}{Observable } & ${R_K}_{[1.1-6.0]}$ 
& ${R_{K^*}}_{[0.045-1.1]}$ & ${R_{K^*}}_{[1.1-6.0]}$& ${P'_{5}}_{[4.0-6.0]}$ & $R_D^{ratio}$ & $R_{D^*}^{ratio}$ &$\frac{B_K^{SM+NP}}{B_K^{SM}}$ & $\frac{B_{K^*}^{SM+NP}}{B_{K^*}^{SM}}$   \\ \hline
\multicolumn{4}{|c|}{Measurement }  &$0.846 \pm 0.062$ 
&$0.66 \pm 0.09$ &$0.69 \pm 0.10$ & $-0.30 \pm 0.16 $&  $1.10 \pm 0.10$ &$1.16 \pm 0.06$  &$\leq 4.0$&$\leq 2.9$   \\ \hline
\multicolumn{4}{|c|}{Standard Model }  &1.0 & 0.93 & 0.99&-0.82 & 1.0 & 1.0& &  \\ \hline
$\theta_{\mu\tau}(rad.)$ & $\theta_{bs}(rad.)$&$g_{U_1}^{33}$ &$g_{U_3}^{33}$   & \multicolumn{8}{|c|}{$U_1 + U_3$}   \\ \hline
$0.31 \pm 0.83 $& $0.05 \pm 0.05$&$-0.69 \pm 0.76$&$0.93\pm 0.72$ & 0.76 & 0.86 & 0.77 & -0.70 & 1.14 & 1.14 & 934.59 & 934.59 \\ \hline
$\theta_{\mu\tau}(rad.)$ & $\theta_{bs}(rad.)$&$g_{U_1}^{33}$ &$g_{S_3}^{33}$   & \multicolumn{8}{|c|}{$U_1 + S_3$}   \\ \hline
$0.64 \pm 0.90$&$1.50 \pm 1.54$&$0.930 \pm 0.002$& $0.928 \pm 0.002$ &  0.74&0.86 &0.75 &-0.69 &1.35 &1.35 & 132.15 & 132.15\\ \hline
$\theta_{\mu\tau}(rad.)$ & $\theta_{bs}(rad.)$&$g_{U_3}^{33}$ &$g_{S_3}^{33}$   & \multicolumn{8}{|c|}{$U_3 + S_3$}   \\ \hline
$0.12 \pm 0.12$& $0.10 \pm 0.11$ &$1.06 \pm 0.81$ &$0.62 \pm 1.05$ & 0.76 & 0.86 & 0.77 & -0.70 & 1.14 & 1.14 & 221.77 & 221.77 \\ \hline
\end{tabular}
}
\caption{Fit results for  $(U_1 + U_3)$, $(U_1 + S_3)$ and $(U_3 + S_3)$ combinations of two LQ particles.}
\label{tab:two_lq}
\end{table}

Finally, we try a combination of two LQ particles taking the LQ couplings to be a free parameter. These results are shown in Table \ref{tab:two_lq}. Clearly, $b\to s\nu \bar \nu$ is a big challenge for  all combinations.

\section{\bf Conclusions}
The  measurements of $R_{K^{(*)}}$  by the LHCb collaboration has 
reinforced the earlier hints of lepton universality violation 
observed in $R_{D^{(*)}}$.  In this work we look for simultaneous explanations of 
these measurements in VB, $U_1$,  $S_3$ and  $U_3$ models. Here we assume a coupling only to the third generation in the gauge basis. Performing `a global fit' to all relevant data, we find that 
 the vector boson model violates the upper bound on the branching ratio of $\tau \to 3\mu$ and hence is inconsistent with the present data.  The $U_1$ LQ model can not accommodate the $R_{K^{(*)}}$ and $R_{D^{(*)}}$
anomalies. This is evident from the fit as well as from the allowed regions which do not overlap  even at 2$\sigma$. We also find that, 
with considered structure of the mixing in this work, the $Br(\tau \to e \nu \bar \nu)$ which arises due to RGE effects, does not put constraint on this model. Further, we find that the $S_3$ and $U_3$ LQ models are highly constrained by the $b \to s \nu \bar \nu$ data.

\section*{Acknowledgment}
We would like to thank Amol Dighe, David London and Uma Sankar for useful suggestions and discussions.

\end{document}